\documentclass[envcountsect,envcountsame]{llncs}
\usepackage{makeidx}
\usepackage{amsfonts}
\usepackage{amsmath}
\usepackage{amssymb}
\usepackage{amscd}
\usepackage[dvips]{graphicx}
\usepackage{algorithmic}
\usepackage{algorithm}
\usepackage{comment}
\usepackage{color}
\usepackage{array,amssymb,amsmath,amsbsy}
\usepackage{enumerate}
\usepackage[bottom]{footmisc}
\usepackage[all,cmtip]{xy}
\usepackage{hyperref}

\makeatletter
\renewcommand*\l@author[2]{}
\renewcommand*\l@title[2]{}
\makeatletter
\newcommand{\nocontentsline}[3]{}
\newcommand{\tocless}[2]{\bgroup\let\addcontentsline=\nocontentsline#1{#2}\egroup}

\newcommand{\heading}[1]{\smallskip\par\noindent{\bf #1}}

\def\computationproblem#1#2#3{% {problem_name}{input}{output}
    \begin{center}
    \begin{tabular}{rp{10cm}}
    {\bf Problem:\enspace}&#1\\
    {\bf Input:\enspace}&#2\\
    {\bf Output:\enspace}&#3\\
    \end{tabular}
    \end{center}
}

  \def\calC{{\cal C}}

 \def\calR{{\cal R}}

\def\cNP{\hbox{\rm \sffamily NP}}

\def\eps{\varepsilon}

\def\RepExt{{\textsc{RepExt}}}
\def\Partition{{\textsc{Partition}}}

\def\int{\hbox{\rm \sffamily INT}}

\def\uca{\hbox{\rm \sffamily UNIT CA}}

\def\eps{\varepsilon}
\spnewtheorem*{observation*}{Observation}{\bfseries}{\rmfamily}
\spnewtheorem*{open}{Open Problem}{\bfseries}{\rmfamily}
\newcounter{lth}
\title{Extending Partial Representations of Unit Circular-arc Graphs}
\author{Peter Zeman}

\institute{Department of Applied Mathematics, Faculty of Mathematics and
Physics,\\ Charles University, Czech
	Republic. E-mail: \texttt{zeman@kam.mff.cuni.cz}.}

\begin{document}
\maketitle

\begin{abstract}
The partial representation extension problem, introduced by Klav\'{i}k et al.
(2011), generalizes the recognition problem. In this short note we show that
this problem is $\cNP$-complete for unit circular-arc graphs.
\end{abstract}

\section{Introduction}

An intersection representation $\calR$ of a graph $G$ is a collection of sets $\{R_v : v \in V(G)\}$
such that $R_u \cap R_v \neq \emptyset$ if and only if $uv \in E(G)$.

\begin{figure}[b]
\centering
\includegraphics[scale=0.8]{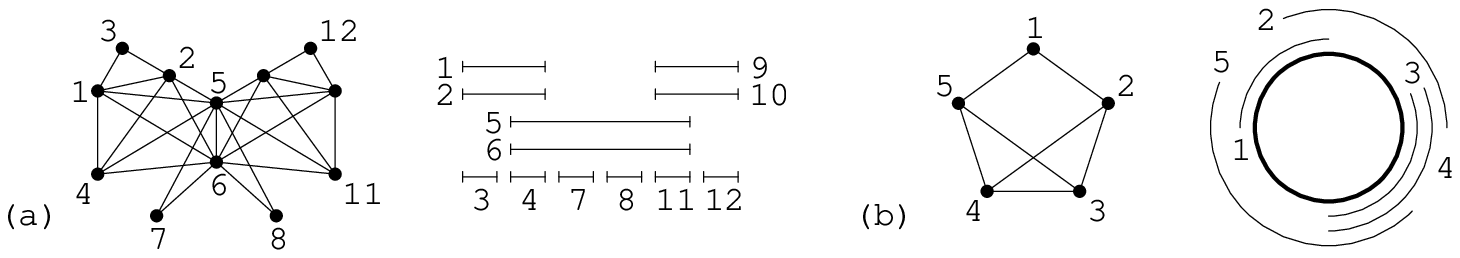}
\caption{(a) An interval graph and one of its interval representations. (b) A
circular-arc graph and one of its representations.}
\label{fig:int_ca_ex}
\end{figure}

\heading{Interval Graphs.}
One of the most studied and well understood classes of intersection graphs are \emph{interval
graphs} (\int). In an \emph{interval representation} of a graph, each set $R_v$ is a closed interval of the
real line. A graph is an interval graph if it has an interval representation; see
Fig.~\ref{fig:int_ca_ex}a.
%We denotethe class of all interval graphs by $\int$.

\heading{Circular-arc Graphs.}
In a \emph{circular-arc representation}, the sets $R_v$ are arcs of a circle; see
Fig~\ref{fig:int_ca_ex}b. 

\heading{Structure of All Representations.}
Despite the fact that circular-arc graphs are a straightforward generalization of interval graphs,
the structure of their representations is much less understood. To understand the structure of all
interval representations, the key result is the following.

\begin{theorem}[Fulkerson and Gross~\cite{maximal_cliques}]
\label{thm:fulkerson_gross}
A graph $G$ is an interval graph if and only if there exists a linear ordering $\preceq$ of its
maximal cliques such that for every vertex $v$, the maximal cliques containing $v$ appear
consecutively in $\preceq$.
\end{theorem}

Booth and Lueker used Theorem~\ref{thm:fulkerson_gross} and PQ-trees~\cite{PQ_trees} to recognize
interval graphs graphs in linear time.  Moreover, a PQ-tree of an interval graph captures all
possible orderings $\preceq$ of the maximal cliques, i.e., it stores every possible representation
of the interval graph.

For circular-arc graphs, the situation is much more complicated. The main difference is that, unlike
a circular-arc representation, an interval representation satisfies the Helly property: if every two
intervals in a set have a nonempty intersection, then the whole set has a non-empty intersection. In
particular, this means the maximal cliques of interval graphs can be associated to unique points of
the line. The number of maximal cliques in an interval graph linear in the number of vertices.
However, a circular-arc representation does not necessarily satisfy Helly property and the number of
maximal cliques can be exponential. The complete bipartite graph $K_{n,n}$ without a matching is an
example of that. It is not clear whether there exits a way to efficiently capture the structure off
all representations of a circular-arc graph. 

\heading{Partial Representation Extension Problem.}
This problem naturally generalizes the recognition problem. For a class of graphs $\calC$, the input
consists of a graph $G$ and a \emph{partial representation $\calR'$} which is a representation of
some induced subgraph $G'$ of $G$. The question is to decide whether there exists a representation
$\calR$ of $G$ that \emph{extends} $\calR'$, i.e., $R_u = R_u'$, for every $u \in V(G')$.  Note that
in the case of recognition, the partial representation $\calR'$ is empty.

\computationproblem
{Partial representation extension -- $\RepExt(\calC)$}
{A graph $G$ and a partial representation $\calR'$.}
{Is there a representation $\calR$ of $G$ extending $\calR'$?}

In recognition it suffices, for a given graph $G$, to construct a single representation of $G$.
However, in partial representation extension, one typically needs a way to store all possible
representations of $G$ efficiently. Then we can efficiently find a representation $\calR$ that
extends $\calR'$. For example, Klav\'{i}k et. al.~\cite{KKV11} used PQ-trees to solve
$\RepExt(\int)$ in linear time.

In the past few years a lot of work was done involving the partial representation extension problem.
This includes circle graphs~\cite{chaplick2013extending}, function and permutation
graphs~\cite{klavik2012extending}, unit and proper interval graphs~\cite{KKORSSV16}, and visibility
representations~\cite{chaplick2015partial}. All of those papers use an efficient way to store all
possible representations and give polynomial-time algorithms for the partial representation
extension problem. For chordal graphs~\cite{KKOS15} and contact representations of planar
graphs~\cite{chaplick2014contact}, the partial representation extension problem is hard.

%Classes for which Partial representation extension problem was considered
%include interval graphs \cite{KKORSSV14}, circle graphs \cite{CFK13},
%permutation and function graphs \cite{KKKW12} and general chordal graphs
%\cite{KKOS15}.

%Clearly, \RepExt\ is a generalization of recognition problem since we can view
%the recognition as if we are given graph $G$ and an empty representation only.
%Extending some pre-drawn representation proves to be more complex task since
%existing representation of an induced subgraph yields more constraints to
%building representation of $G$.

We argued that studying the partial representation extension problem of a given class of graphs is
closely related to understanding the structure of all representations. The problem $\RepExt$ can be
typically solved in polynomial time if we can store all possible representations efficiently.

\heading{Unit Circular-arc Graphs.}
Circular-arc graphs with an intersection representation in which every arc has a unit length are
called \emph{unit circular-arc graphs} (\uca). An example of a circular-arc graph that is not unit
is the complete bipartite graph $K_{1,3}$.

\begin{theorem}\label{thm:uca_npc}
The problem $\RepExt(\uca)$ is $\cNP$-complete.
\end{theorem}

Note that for unit interval graphs (defined analogously) $\RepExt$ can be solved in polynomial
time~\cite{KKORSSV16}.

\section{Proof of The Main Result}

We prove Theorem~\ref{thm:uca_npc}. The problem $\RepExt(\uca)$ is clearly in $\cNP$. We show a
reduction from a known $\cNP$-complete problem called $3$-$\Partition$~\cite{garey1975complexity}.
The input of $3$-$\Partition$ consists of positive integers $k$, $M$, and $A_1, \dots, A_{3k}$ such
that $M/4 < A_i < M/2$, for each $A_i$, and $\sum A_i = kM.$ The problem asks whether it is possible
to partition $A_i$'s into $k$ triples such that the sets $A_i$ belonging to the same triple sum up
to exactly $M$.  (Note that the size constraints on $A_i$'s ensure that every subset that sums
exactly to $M$, is a triplet.)

\begin{proof}[Theorem~\ref{thm:uca_npc}]
For a given instance of $3$-$\Partition$, we construct a unit circular-arc graph $G$ and its partial
representation $\calR'$. For technical reasons, we assume that $M \geq 8$.

Let $P_{2\ell}$ be a path of length $2\ell$. There exists a unit circular-arc representation
$P_{2\ell}$ such that it spans $\ell + \eps$ units, for some $\eps > 0$. To see this, note that
$P_{2\ell}$ has two independent sets of size $\ell$ and each of this independent sets needs at least
$\ell + \eps$. Let $a, b, c$ be positive integers such that $a + b + c = M$. It follows that the
disjoint union of $P_{2a}$, $P_{2b}$, and $P_{2c}$ has a representation such that it spans $M +
\eps$ units, for some $\eps > 0$, and therefore, it can be fit into $M+1$ units; see
Fig.~\ref{fig:unit_proof}.

\begin{figure}[b]
\centering
\includegraphics[scale=0.8]{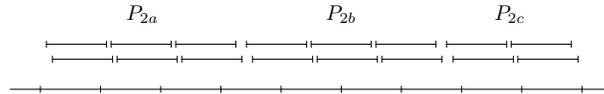}
\caption{A representation of the disjoint union of $P_{2a}$, $P_{2b}$, and $P_{2c}$ fits into $M+1$
units. Here, $M = 8$, $a = 3$, $b = 3$, and $c = 2$.}
\label{fig:unit_proof}
\end{figure}

Let $x_0, \dots, x_{k(M+2)-1}$ be points of the circle that divide it into $k(M+2)$ equal parts,
i.e., vertices of a regular $k(M+2)$-gon.  The graph $G$ is a disconnected graph consisting of $4k$
connected components.  For each $A_i$, we take the path $P_{2A_i}$. We further add an isolated
vertex $v_j$, for $j = 0, \dots, k-1$. The partial representation $\calR'$ is the collection
$\{R_{v_j} : j = 0,\dots, k-1\}$, where $R_{v_j}$ is the arc of the circle from $x_{j(M+2)}$ to
$x_{j(M+2)+1}$ in the clockwise direction.

The pre-drawn arcs $R_{v_0}, \dots, R_{v_{k-1}}$ split the circle into $k$ gaps, where each gap has
exactly $M+1$ units. By the discussion above, if the $A_i$'s can be partitioned into $k$ triples
such that each triple sums to $M$, then a representation of the disjoint union of the paths
corresponding to a triple can be placed in one of the $k$ gaps. If the partial representation
$\calR'$ can be extended, then we a have partition of the $A_i$'s into $k$ triples such that each
triple sums to $M$.
\qed
\end{proof}

%\nocite{*}
\bibliographystyle{plain}
\bibliography{extending_unit}
\end{document}